\documentclass[a4paper, 12pt]{article}
\usepackage[english]{babel}
\usepackage[cp1250]{inputenc}
\usepackage[T1]{fontenc}
\usepackage{epsfig}

\begin{document}
\title{Symmetry breaking transition and appearance of compactons in
a mechanical system
\thanks{Paper supported in part by ESF ``COSLAB'' Programme  }}
\author{H. Arod\'z \\
Marian Smoluchowski Institute of Physics, Jagiellonian University, \\
Reymonta 4, 30-059 Cracow, Poland}
\date{$\;\;\;$}
\maketitle

\vspace*{2cm}

\begin{abstract}
Recently we have described a mechanical system which exhibits spontaneous breaking of $Z_2$ symmetry and related
topological kinks called compactons. The corresponding field potential is not differentiable at its global
minima. Therefore, standard derivation of dispersion relation $\omega(k)$ for small perturbations around the
ground state can not be applied. In the present paper we obtain the dispersion relation. It turns out that
evolution equation remains nonlinear even for arbitrarily small perturbations. The shape of the resulting
running wave is piecewise combined from $\pm cosh$ functions.  We also analyze dynamics of the symmetry breaking
transition. It turns out that the number of produced compacton-anticompacton pairs strongly depends on the form
of initial perturbation of the unstable former ground state.
\end{abstract}

\vspace*{2cm}
\noindent
Preprint TPJU-8/2003 \\
PACS numbers: 03.50.Kk, 11.10.Lm

\pagebreak

\section{Introduction}

During the recent decades it has become clear that spontaneous symmetry breaking and related with it topological
defects play important role in physics of superfluid Helium, superconductors, liquid crystals, hadronic matter,
cosmic objects like neutron stars or hypothetical cosmic strings, and many other systems. Reviews and references
can be found in, e.g., \cite{1,2}. Static properties of the topological defects are understood relatively well.
Dynamics of them is much harder to unravel. Interactions and production of the defects in general involve many
modes of underlying fields (or order parameters), and non-trivial time-dependent solutions of nonlinear field
equations have to be investigated.

In the present paper we discuss time-dependent solutions in the effective model introduced in \cite{3}.  The
model is distinguished by several attractive features. First, the pertinent field equations are very simple.
Second, the model has clear relation with a classical system of macroscopic pendulums, which is easy to
visualize, and even to build. This is in contrast with many other effective models with topological defects
where their relation with the pertinent physical system is not transparent -- this is especially true in the
case of classical Ginzburg - Landau type effective models for quantum fluids. Finally, the topological defects
in our model have only contact interactions  -- they do not interact until they touch each other. The reason is
that solutions describing the defects exactly coincide with vacuum solutions outside certain finite region in
the space, hence the common exponential tails are absent here. Such  defects are called compactons\footnote{This
name has been proposed in connection with solitonic solutions of modified Korteveg - deVries equations, see,
e.g., \cite{4,5,6}.}.

The Lagrangian obtained in \cite{3} has the form
\begin{equation}
L= \frac{1}{2} (\partial_{\tau}\phi)^2 - \frac{1}{2} (\partial_{\xi}\phi)^2
- V(\phi),
\end{equation}
where $\xi$ is a dimensionless coordinate along a straight-line, $\phi = \phi (\xi, \tau) $ is the scalar field
which in general can depend on $\xi $ and on a dimensionless time $\tau$. The field potential $V(\phi)$ is given
by the following formula
\begin{equation}
V(\phi) = \left\{
\begin{array}{lcl}
\cos\phi -1  &  {\rm for} &  | \phi| \leq \phi_0 \\
\infty  &  {\rm for}  &   |\phi| > \phi_0,
\end{array}
\right.
\end{equation}
where  $\phi_0$ is a constant, $\pi >\phi_0 >0$.  The potential has two
degenerate minima at $\phi= \pm \phi_0$, see Fig. 1.  \\
%\psdraft
\begin{center}
\hspace*{0.5cm}  \epsfig{file=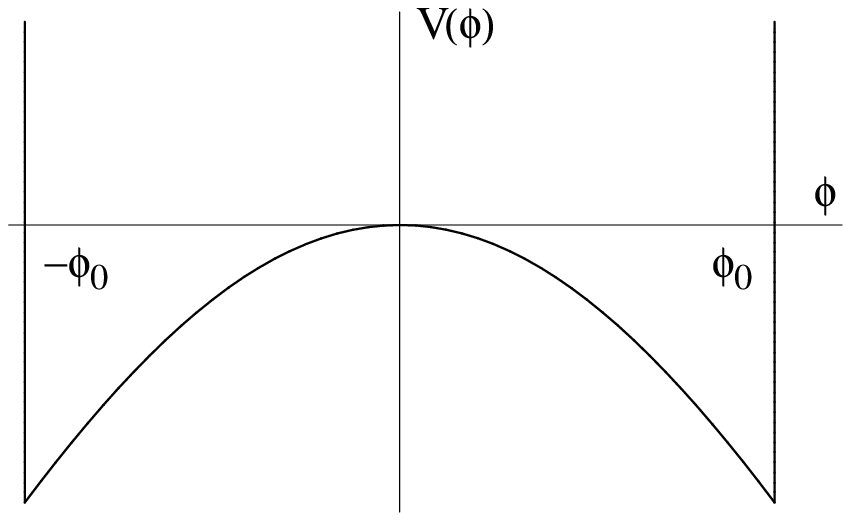}
\end{center}
\begin{center}
Fig.1. The potential $V(\phi)$.
\end{center}

The variable $\xi$ is restricted to the interval $[- \xi_N, \xi_N].$ By assumption, the field $\phi$ obeys the
following boundary conditions
\begin{equation}
\phi(\pm \xi_N, \tau) = 0.
\end{equation}
Because of the infinite potential barrier, modulus of the field $\phi$ cannot
exceed $\phi_0,$
\begin{equation}
-\phi_0 \leq \phi \leq \phi_0.
\end{equation}
Scalar fields with values restricted to certain finite interval appear also in a completely different context of
brane models, \cite{7}.

As discussed in detail in \cite{3}, the effective model defined by formulas (1-4) describes a system of $2N+1$
pendulums attached to a rectilinear wire at the constant distance $a$ between them. Each pendulum has the arm of
length $R$, and  mass $m$ on the free end which can move only in a plane perpendicular to the wire. $\phi$ is
the angle between the arm and the vertical upward direction. The pendulums are firmly attached to the wire, so
that $\phi$ is also the angle by which the wire is twisted. It is assumed that the wire is elastic with respect
to the twists, $\kappa$ is the corresponding torsional elasticity coefficient. The angle $\phi$ is restricted to
the interval $[-\phi_0, \phi_0]$  by two stiff rods placed symmetrically on both sides of the wire, parallel to
it and at the appropriate distance. The boundary conditions (3) mean that the two outermost pendulums are kept
at rest in the upward position by an external force. The dimensionless variables $\xi, \tau$ are related to the
physical position coordinate $x$ (along the wire) and time $t$ by the formulas
\begin{equation}
\xi = \sqrt{\frac{mgR}{\kappa a}} x,
 \;\;\; \tau = \sqrt{\frac{g}{R}} t.
\end{equation}
The two outermost pendulums are located at $x= \pm Na,$ hence
\begin{equation}
\xi_N = \sqrt{\frac{mgRa}{\kappa}} N.
\end{equation}

The system of pendulums as described above, but without the two rods, is well-known in connection with
sine-Gordon equation and its solitonic solutions \cite{8}. In that case the space of  values of $\phi(\xi,
\tau)$ is a circle $S^1$, and the solitons appear because of nontrivial topology of $S^1$. In our case, instead
of $S^1$ we have the topologically trivial interval $[-\phi_0, \phi_0].$ The presence of compactons is due to
spontaneous breaking of the $Z_2$ symmetry
\begin{equation}
\phi(\xi, \tau) \rightarrow - \phi(\xi, \tau).
\end{equation}
It takes place when $\xi_N$ becomes large enough, see Section 3.  Yet another system of coupled pendulums is
described in \cite{9}. It differs from the sine-Gordon or our system by the presence of a term which represents
nonlinear elasticity of the wire, and from our system also by absence of the rods. Thanks to the nonlinear
elasticity the solitons are compactons. They essentially differ from the ones discussed by us.

In \cite{3} we have discussed in detail the relation between the system of pendulums and the field-theoretic
model with Lagrangian (1), as well as basic properties of compactons: their shape and energy. In the present
paper we investigate waves propagating in the ground state with broken $Z_2$ symmetry, and we make a preliminary
study of production of compactons and anti-compactons during the $Z_2$ symmetry breaking transition. Such
transition can be triggered by, for example, decreasing the elasticity coefficient $\kappa$ or by increasing the
length $R$ of the arm. Both possibilities seem technically feasible. The first topic is discussed in Section 2.
It is interesting because the standard procedure, based on linearization of field equations, does not work here
-- potential (2) is not smooth at the absolute minima. It turns out that even in the limit of small amplitudes
the pertinent evolution equation remains nonlinear.  As for the second topic, it is related to the broad subject
of dynamics of rapid phase transitions and production of topological defects \cite{10, 11}. We show in Section 3
that thanks to extraordinary simplicity of our model one can calculate time evolution of the system during the
symmetry breaking transition. In particular, one can see how many topological defects are produced.

\section{Small perturbations of the ground state}

In this section we consider an infinite chain of pendulums, $\xi_N \rightarrow \infty,$ and we abandon the
boundary conditions (3). The main reason for this step is that we would like to study  effects which are due to
the infinite potential barrier at $\phi = \pm \phi_0.$ At the ends of the finite chain the barrier is not felt
at all because of condition (3).

The evolution equation which corresponds to Lagrangian (1) has the form
\begin{equation}
\partial_{\tau}^2 \phi - \partial^2_{\xi}\phi - \sin\phi =0
\end{equation}
if
\[
| \phi | < \phi_0.
\]
When $\phi$ reaches $\pm \phi_0$ the pendulums elastically bounce
from the rods,
\begin{equation}
\partial_{\tau}\phi \rightarrow - \partial_{\tau}\phi,
\end{equation}
or just rest on them if the velocity vanishes. Let us emphasize
that Eq.(8) does not hold when $\phi = \phi_0.$

In the present paper we assume that the maximal angle $\phi_0$ is
small and therefore
\begin{equation}
\sin \phi \cong \phi.
\end{equation}
This can be achieved by putting the rods close enough to the pendulums when they are in the upward vertical
position. Then, Eq.(8) is linearized
\begin{equation}
\partial_{\tau}^2 \phi - \partial^2_{\xi}\phi - \phi =0.
\end{equation}

For the infinite chain of pendulums the two degenerate ground states are given by
\begin{equation}
\phi = \pm \phi_0.
\end{equation}
Evolution of small perturbations of the ground state $\phi_0$ is governed by Eq.(11) and the reflection
condition (9). Such perturbations have the form of waves propagating along the chain: the pendulums rise a
little bit above the rod and fall back. We assume that amplitudes of the waves, equal to $\phi_0 - \phi(\xi,
\tau),$ are not too large,
\begin{equation}
\phi_0 - \phi(\xi, \tau) < \phi_0,
\end{equation}
that is that none of the pendulums  reaches the upward vertical position. Condition (13) is of course equivalent
to $\phi(\xi, \tau) >0.$ In order to investigate the waves we use the following trick. Instead of the original
field $\phi(\xi, \tau)$ we consider another one, namely  $\underline{\phi}(\xi, \tau)$ related to
$\phi(\xi,\tau)$  by the formula
\begin{equation}
\phi(\xi, \tau) = \left\{
\begin{array}{lcl}
\underline{\phi}(\xi, \tau) & \mbox{if} &     \underline{\phi}(\xi, \tau)\leq \phi_0 \\
2 \phi_0 - \underline{\phi}(\xi, \tau) & \mbox{if} &  \underline{\phi}(\xi, \tau) \geq \phi_0.
\end{array}\right.
\end{equation}
Condition (13) means that
\begin{equation}
 0 < \underline{\phi} < 2 \phi_0.
\end{equation}
Equation (11) is equivalent to the following evolution equation for $\underline{\phi}$
\begin{equation}
(\partial_{\tau}^2 - \partial_{\xi}^2)\underline{\phi} = \left\{
\begin{array}{lcc}
\underline{\phi} & \mbox{if} &  \underline{\phi} < \phi_0, \\
0 & \mbox{if} & \underline{\phi} = \phi_0 \\
\underline{\phi} - 2 \phi_0 & \mbox{if} & \underline{\phi} > \phi_0.
\end{array}
\right.
\end{equation}
The r.h.s. of this equation for $\underline{\phi} \neq \phi_0$  corresponds to the  potential
$\underline{V}(\underline{\phi})$ given by the following formula
\begin{equation}
\underline{V}(\underline{\phi}) = \left\{
\begin{array}{lcl}
- \frac{1}{2} \underline{\phi}^2 & \mbox{for} & 0 < \underline{\phi} \leq \phi_0, \\
- \frac{1}{2} (\underline{\phi}- 2 \phi_0)^2  &\mbox{for} & \phi_0 \leq \underline{\phi} < 2\phi_0.
\end{array}
\right.
\end{equation}
$\underline{V}(\underline{\phi})$ is symmetric with respect to the point $\underline{\phi} = \phi_0$ at which it
reaches its absolute minimum, $\underline{V}(\phi_0) = - \phi_0^2/2,$ see Fig. 2. Equation (16) gives continuous
evolution of $\underline{\phi}$ and of its first derivatives $\partial_{\tau}\underline{\phi}, \;\;
\partial_{\xi}\underline{\phi}$.
The advantage of using $\underline{\phi}$ instead of $\phi$ is that because of relation (14) such continuous
evolution of $\partial_{\tau}\underline{\phi}$ automatically  gives the reflection (9)  of $\partial_{\tau}\phi$
at $ \phi = \phi_0.$

\begin{center}
\hspace*{0.5cm}  \epsfig{file=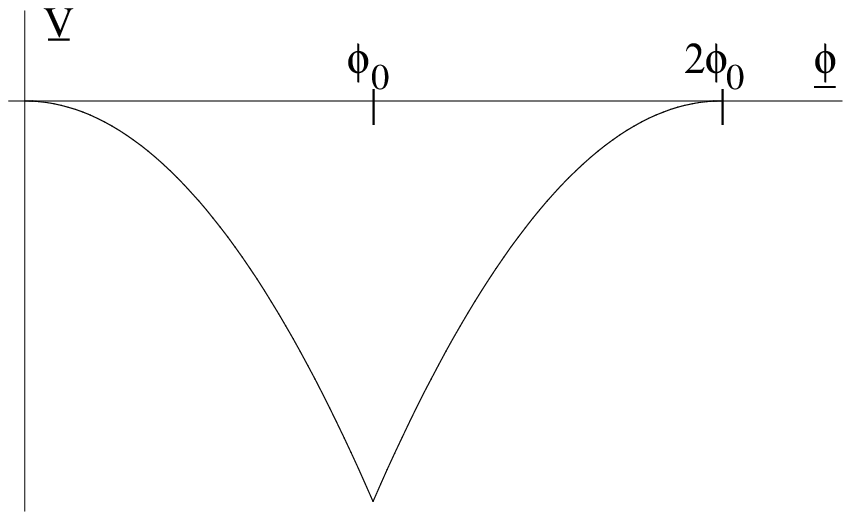}
\end{center}
\begin{center}
Fig.2. The potential $\underline{V}(\underline{\phi}).$
\end{center}

Because we are interested in oscillations of $\underline{\phi}$ around $\phi_0$, we substitute
\[
\underline{\phi} = \phi_0 + \phi_0 \epsilon(\xi, \tau).
\]
Evolution equation for $\epsilon$ follows from Eq.(16),
\begin{equation}
(\partial_{\tau}^2 - \partial_{\xi}^2)\epsilon =  \epsilon  - \mbox{sign}( \epsilon)
\end{equation}
(we assume that $  \mbox{sign}( 0) =0).$   Condition (15) is equivalent to
\begin{equation}
|\epsilon | < 1.
\end{equation}
Solutions of Eq.(18)  describing running periodic waves can be obtained with the help of the Ansatz
\begin{equation}
\epsilon(\xi, \tau) = \epsilon(\zeta), \;\;\; \zeta = \frac{\xi - v\tau}{\sqrt{v^2 -1}}.
\end{equation}
Here $v$ is the phase velocity of the wave. It is assumed that $|v| >1.$ Now equation (18) is reduced to
ordinary differential equation
\begin{equation}
\epsilon^{''} = \epsilon - \mbox{sign}(\epsilon),
\end{equation}
where $^{'}$ denotes $d/d\zeta$. Integration of Eq.(21) gives the relation
\[
\epsilon^{'2} = (1 -|\epsilon|)^2 - (1-\epsilon_0)^2,
\]
where $\epsilon_0 >0$ is the maximal value of $\epsilon$. Hence,
\begin{equation}
\pm \epsilon' = \sqrt{\epsilon_0 - |\epsilon|} \sqrt{2 - \epsilon_0 - |\epsilon|}.
\end{equation}
the sign $+$ or $-$ should be chosen in accordance with the sign of $\epsilon'$ because the r.h.s. of (22) is
nonnegative. We consider Eq.(22) in the regions $\epsilon \geq 0$ and $ \epsilon \leq 0$ separately. Simple
calculations show that if $\pm \epsilon \geq 0$ then $\epsilon = \epsilon_{\pm}$, respectively, where
\begin{equation}
\epsilon_+(\zeta) = 1 - (1 - \epsilon_0) \cosh(\zeta - \zeta_0),
\end{equation}
\begin{equation}
\epsilon_-(\zeta) = -1 + (1-\epsilon_0) \cosh(\zeta - \zeta_1).
\end{equation}
Here $\zeta_0, \zeta_1$ are constants. These two solutions should match each other at $\zeta = \zeta_{1/2}$ such
that $\epsilon_+ =0 = \epsilon_-.$ Thus,
\[
 1 = (1 - \epsilon_0) \cosh(\zeta_{1/2} - \zeta_0),
\]
and
\[
1 = (1 - \epsilon_0) \cosh(\zeta_{1/2} - \zeta_1).
\]
These conditions are satisfied if
\begin{equation}
\zeta_{1/2} = \zeta_0 + \frac{\Lambda}{4}, \;\; \zeta_1 = \zeta_0 + \frac{\Lambda}{2},
\end{equation}
where
\begin{equation}
\Lambda = 4\; \mbox{arcosh}\frac{1}{1-\epsilon_0}.
\end{equation}
It turns out that at such $\zeta_{1/2}$ also the first derivatives of $\epsilon_+, \epsilon_-$ are equal.
$\epsilon_+(\zeta) $ is non-negative in the interval $\zeta \in [ \zeta_0 - \Lambda/4, \zeta_0 + \Lambda/4]$,
and $\epsilon_-(\zeta) $ is non-positive on the interval $\zeta \in [ \zeta_1 - \Lambda/4, \zeta_1 +
\Lambda/4]$. Taken together they give $\epsilon$ in the interval $[\zeta_0 - \frac{\Lambda}{4}, \zeta_0 +
\frac{3}{4} \Lambda ]$, and $\epsilon$ for all other values of $\zeta$ can be obtained by periodic repeating
$\epsilon_+$ and $\epsilon_-$, see Fig.3. Thus, we have obtained the running periodic wave of the $\cosh$ type.

%\psdraft
\begin{center}
\hspace*{0.5cm}  \epsfig{file=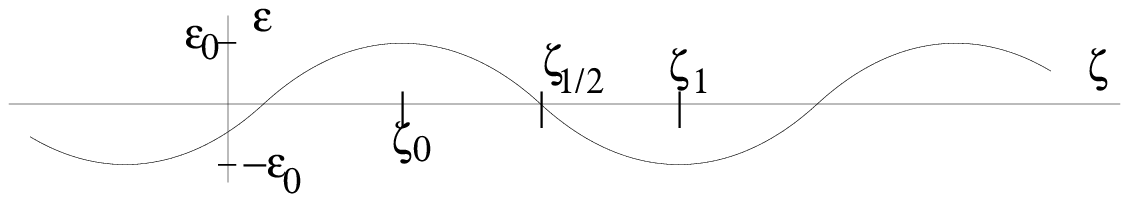}
\end{center}
\begin{center}
Fig.3.  The periodic wave of the cosh type.
\end{center}

It is clear that $\Lambda$ is the wave length for $\epsilon(\zeta)$. The periodicity in $\zeta$ implies
periodicity in $\xi$ with the wave length
\[
\lambda = \Lambda \sqrt{v^2-1},
\]
and in $\tau$ with the period
\[
T= \Lambda \sqrt{1- \frac{1}{v^2}}.
\]
The corresponding wave number $k = 2 \pi/ \lambda$ and the frequency $\omega = 2 \pi / T$ obey the relation
\begin{equation}
\omega^2 - k^2 = \frac{4\pi^2}{\Lambda^2}.
\end{equation}
The group velocity of this wave
\[
v_g = \frac{\partial \omega}{\partial k} = \frac{k}{\omega} = \frac{1}{v}
\]
is smaller that 1. Because the r.h.s. of formula (27) does not vanish, the waves are massive. For fixed wave
number $k$ the group velocity increases with $\Lambda$. According to formula (26) $\Lambda$ increases with
$\epsilon_0.$ Therefore, the waves with larger amplitude are faster. This dependence on the amplitude appears
because Eq.(21) is nonlinear. Due to the discontinuous character of the $\mbox{sign}$ function it can not be
linearized even for arbitrarily small $\epsilon$ --  in our system even arbitrarily small oscillations around
the ground state are nonlinear.

\section{The $Z_2$ symmetry breaking transition}

In this Section we consider the finite chain of pendulums, $\xi_N < \infty.$ $ \phi$ obeys the boundary
conditions (3). We again assume that $\phi_0$ is small enough to justify the approximation (10). In order to
simplify the model further, we now assume that the dynamics of pendulums is dissipative, that is that the
pertinent evolution equation has the form
\begin{equation}
\partial_{\tau} \phi - \partial^2_{\xi}\phi - \phi =0,
\end{equation}
instead of (11) (if $|\phi| < \phi_0.)$ Physically this can be achieved by putting the chain of pendulums in a
viscous liquid. In the present case the second of formulas (6), relating the $\tau$ variable with the physical
time $t,$ is replaced by
\[
\tau = \frac{mg}{\gamma R} t,
\]
where $\gamma $ is a constant coefficient characterizing the friction. The main advantage of such over-damped
system is that by  quickly loosing its energy it rapidly evolves towards a stable state.

If the chain of pendulums is long enough, that is if $\xi_N \geq \pi/2$ as shown below, there are two degenerate
ground states given by
\[
\phi = \pm \phi_v,
\]
where
\begin{equation}
\phi_v = \left\{
\begin{array}{lcl}
 \phi_0 \sin (\xi + \xi_N)  & \mbox{for} &  \xi \in [-\xi_N, -\xi_N +
\frac{\pi}{2}],  \\
 \phi_0 & \mbox{for} & \xi \in [-\xi_N + \frac{\pi}{2}, \xi_N -
\frac{\pi}{2}],  \\
 \phi_0 \sin (\xi_N - \xi) & \mbox{for} & \xi \in [\xi_N - \frac{\pi}{2}, \xi_N ].
\end{array}
\right.
\end{equation}
Such $\phi_v$ obeys the boundary condition (3). The sinus functions are solutions of Eq.(28) -- they describe
gradually tilted pendulums. For the intermediate values of $\xi$, that is when $-\xi_N + \pi/2 \leq \xi \leq
\xi_N - \pi/2,  $ the pendulums just rest on one rod. For the chains shorter than $\pi$ formula for $\phi_v$
given above can not be accepted. Below we show that in this case the stable ground state is given by
\begin{equation}
\phi_s(\xi) = 0.
\end{equation}
This state is symmetric with respect to the $Z_2$ transformation (5).

The static compacton is obtained from the ground state $-\phi_v$ by flipping pendulums from one rod to the other
one  at the intermediate values of $\xi$. This flip is given by the static solution of Eq.(28),
\begin{equation}
\phi = \phi_0 \sin(\xi - \xi_0)
\end{equation}
in the interval $ \xi \in ( \xi_0 - \pi/2, \xi_0 + \pi/2).$ The compacton is located at $\xi_0$ and it has the
width $\pi$. Notice that the sinus functions present in formula (29) can be regarded as half-compacton and
half-anticompacton sitting at the ends of the wire. If the compacton is present, at the ends we have two
half-anticompactons.  The compacton does not interact with these half-anticompactons  if it is located far
enough from the ends, namely if  $ -\xi_N + \pi < \xi_0 < \xi_N - \pi. $

Solutions of Eq.(28) can be written in the form of decomposition into eigenmodes
\begin{equation}
\phi(\xi, \tau) = \sum_{n=1}^{\infty} f_n(\tau) \sin\frac{\pi n}{2\xi_N}(\xi + \xi_N),
\end{equation}
where
\begin{equation}
f_n(\tau) = f_n(0) \exp(\lambda_n\tau),
\end{equation}
\begin{equation}
\lambda_n = 1 - \frac{\pi^2 n^2}{4 \xi_N^2}.
\end{equation}
Let us recall that $\phi(\xi,\tau)$ evolves according to equation (28) until a pendulum touches one of the rods.
Therefore,  solution (32) is physically relevant until $|\phi| = \phi_0$ at certain $\xi$ and $\tau$. At later
times the description of the evolution of the system has to take into account the reflection condition (9).
Obviously, formulas (32-34) correctly describe evolution of small deviations from the $Z_2$ symmetric state
$\phi_s =0.$ In the case when this state is unstable, these formulas are relevant if $\tau$ is in certain finite
interval $[0, \tau_0].$ If that state is stable, $ \lambda_n <0$, the formulas describe exponential
disappearance of small perturbations when $\tau \rightarrow \infty.$ Of course, in any case we have to take the
coefficients $f_n(0)$ small enough.

Formulas (33-34) show that the modes with $n> 2\xi_N / \pi$ exponentially disappear, while the modes with $ n <
2 \xi_N/\pi$ exponentially grow. Because the smallest value of $n$ is equal to 1, all modes exponentially
disappear if $\xi_N < \pi/2$. This means that the $Z_2$ symmetric solution $\phi_s =0$ is stable. In this case
$\xi_N$ is too small to allow for appearance of the half-compacton + half-anticompacton pair at the ends of the
chain.

Let us remark here that such clear cut division into the two possibilities is an artifact of the approximation
(10). As shown in \cite{3}, in the case the compacton  is obtained from Eq.(8) without this approximation, its
width depends on $\phi_0$ and it approaches $\pi$ when $\phi_0 \rightarrow 0$ (the difference vanishes as
$\phi_0^2$).  Therefore, in the real system $ \xi_N $ larger than $\pi/2$ still can be too small to accomodate
the half-compactons at the ends. In such intermediate case certain unstable modes $f_n(\tau), \; 1 \leq n \leq
n_0,$ are present but in the final static and stable configuration the pendulums are tilted towards one of the
rods but do not reach it.

The value of $\xi_N$ can be regulated by changing $\kappa$. It turns out that evolution of the system  can be
investigated analytically also when $\kappa$ is time-dependent,  $\kappa = \kappa(\tau).$ This is particularly
interesting when $\kappa$ is changed from a large value $\kappa_i > \kappa_c$ to a small value $\kappa_f <
\kappa_c$, where
\[
\kappa_c = \frac{4mgRaN^2}{\pi^2}
\]
is the critical value of the elasticity coefficient at which the first unstable mode appears and the symmetric
state looses its stability.

First, we return to the physical position variable $x$ because it is independent of $\kappa$, see the first of
formulas (6) -- equation (28) is written in the following form
\begin{equation}
\partial_{\tau} \phi - \frac{\kappa a}{mgR} \partial^2_{x}\phi - \phi =0,
\end{equation}
where $x \in [-Na, Na].$ Let us write $\phi(x, \tau)$ in the form of Fourier series
\begin{equation}
\phi(x, \tau) = \sum_{n=1}^{\infty} h_n(\tau) \sin\frac{\pi n}{2Na}(x+Na).
\end{equation}
Equation (35) implies that
\begin{equation}
\dot{h}_n(\tau) = \lambda_n h_n(\tau),
\end{equation}
where
\begin{equation}
\lambda_n = 1 - \frac{\pi^2 n^2 \kappa}{4mgRaN^2} = 1- \frac{ n^2 \kappa}{\kappa_c},
\end{equation}
the dot denotes $d/d\tau.$  Equations (37) can be integrated also when $\lambda_n = \lambda_n(\tau):$
\begin{equation}
h_n(\tau) = h_n(0) \exp\left(\int_0^{\tau}\lambda_n(\sigma) d\sigma \right).
\end{equation}

Let consider in detail the transition which begins at the moment $\tau =0,$ and such that
\begin{equation}
\kappa(\tau) = \left\{
\begin{array}{ccc}
\kappa_i - \dot{K} \tau & \mbox{if} & 0 \leq \tau \leq \tau_f \\
\kappa_f & \mbox{if} & \tau \geq \tau_f,
\end{array}
\right.
\end{equation}
where $\dot{K} >0$ is the constant rate of change of the elasticity coefficient $\kappa$. Hence,
\[ \tau_f =
\frac{\kappa_i - \kappa_f}{\dot{K}}.
\]
In this case formula (39) gives
\begin{equation}
h_n(\tau) = h_n(0) \exp\left( \tau - \frac{\kappa_i}{\kappa_c} n^2 \tau + \frac{\dot{K}}{2\kappa_c} n^2
\tau^2\right)
\end{equation}
when $\tau \leq \tau_f,$ while at later times
\begin{equation}
h_n(\tau) = h_n(0) \exp\left( - \frac{(\kappa_i - \kappa_f)^2 n^2}{2\kappa_c \dot{K}}\right) \exp\left( \tau -
 \frac{\kappa_f}{\kappa_c} n^2 \tau \right).
\end{equation}

Formula (41) implies that $n$-th mode begins to grow at $\tau = \tau_n,$ where
\begin{equation}
\tau_n = \frac{2\kappa_i}{\dot{K}} \left(1- \frac{\kappa_c}{\kappa_i n^2}\right).
\end{equation}
However, one has to add the restriction $\tau_n \leq \tau_f$ because for larger times formula (41) is not
relevant. This restriction implies that in fact only modes with $n \leq n_{f1}$ become unstable, where
\begin{equation}
n_{f1}^2 = \frac{2 \kappa_c}{\kappa_i + \kappa_f}.
\end{equation}
Because $\kappa_f >0$ and $\kappa_i > \kappa_c,$
\[
n_{f1} \leq 1.
\]
The term with $n=1$ in formula (36) describes pendulums which are tilted to one of the rods only -- this mode
does not give $ \phi=0$ at any point except the ends. Therefore, this mode can not lead to appearance of the
kinks.  Hence, right during the transition, ie. when $\tau \leq \tau_f,$ no compactons and anti-compactons are
produced.

During the second stage of evolution of the system, when $\tau \geq \tau_f,$ formula (42) should be used. It
predicts that the number of growing modes remains constant, and that it is bounded from above by $n_{f2},$ where
\[
n_{f2}^2 = \frac{\kappa_c}{\kappa_f}.
\]
It is clear that $n_{f2}> 1,$  and that $n_{f2}$ becomes arbitrarily large if $\kappa_f$ is smaller and smaller.
Below we argue that $n_{f2}$ gives the upper bound on the produced compacton - anticompacton  pairs. Notice that
it does not depend on $\kappa_i$. Also the magnitude of $\dot{K},$ or equivalently $\tau_f$, is not relevant in
this respect.

According to formula (36) each mode evolves independently. Therefore, the presence or absence of a given mode is
determined by the initial data specifying $h_n(0)$. If only a single growing mode $h_n$  is present, its
amplitude grows monotonically until the pendulums touch one of the rods. If the friction is very large, one may
expect that they will not be able to bounce back over the energy barrier at $\phi=0$ and to fall on the other
rod. Therefore, in such over-damped case the total number of produced compactons and anti-compactons is equal to
$n-1.$ This number does not include the half-(anti)compactons which appear at the ends of the chain. The number
of compactons may differ by 1 from the number of anti-compactons -- in such a case the total topological charge
is made equal to zero (which is its initial value) by the two halves at the ends. For example, if a single
compacton is present, at both ends there are two half-anticompactons. Thus, in the over-damped case the total
number of produced compactons and anti-compactons is not larger that $n_{f2} -1$, and it does not depend neither
on the initial value of $\kappa$ nor on the rate  $\dot{K}$ of the transition. The actual number of produced
kinks is determined by the initial data $h_n(0)$.

Formula (41) shows that the value  of $h_n(\tau_f)$ increases with increasing value of the rate  $\dot{K}$ of
the transition if $n< n_c$, and decreases if $n >n_c$, where
\[
n_c^2 = \frac{2\kappa_f}{\kappa_f + \kappa_i} n_{f2}^2.
\]
It is clear that $n_c < n_{f2}.$
 Because evolution of the modes at times larger than $\tau_f$ does not depend on $\dot{K}$, this
means that the unstable modes with $n < n_c$  have larger amplitudes if the transition is rapid. Then, these
modes reach the angles $\pm \phi_0$ earlier than in the case of a slower transition with the same initial
perturbation. The opposite happens for the modes with $n > n_c.$ Thus, the magnitude of $\dot{K}$ has influence
on the time of arrival at the final state after the transition.

\section{Remarks}

\noindent 1. One can develop another approach to studying the waves propagating in the ground state $\phi_0$
(Section 2) by first smoothing the potential $V(\phi)$ at $\phi = \phi_0$, then applying the standard approach,
and removing the regularization in the end. The essential difficulty  with this approach is that in order to
obtain waves of non-vanishing amplitude after returning to the original potential $V(\phi)$ one has to consider
waves with finite amplitudes in the regularized potential, and such waves probe nonlinearity of the regularized
potential. Therefore, this approach is in fact more cumbersome than the one adopted in Section 2.

\noindent 2. It is tempting to compare our results on production of the kinks  with the well-known
phenomenological predictions for the number of produced topological defects during phase transitions in
condensed matter systems, see \cite{1, 12} for reviews and references, where a strong dependence on the rate of
the transition is expected, and actually seen in various numerical simulations and actual experiments.  On the
other hand, our explicit calculations of that number give results which do not show such dependence. This may be
explained by the fact that we have tackled only the case of very special initial perturbations: the ones with a
single eigen-mode present. Clearly, studies of this aspect of the model should be continued by considering more
generic initial states.

\noindent 3. Another topic which deserves a separate study is interaction of compactons, e.g., their scattering.
Preliminary investigations reveal a rich dynamics which includes emission of radiation and breather-like states.
One more interesting aspect not discussed here is interaction of the waves and the kinks with the boundaries at
$\pm \xi_N$. Recently a study of such interactions in another model has appeared \cite{13}.

\section{Acknowledgements}
I would like to thank Danielle Steer for informing me about the link with the brane models,   Pawe\l{}
W\c{e}grzyn and Zdobys\l{}aw \'Swierczy\'nski for useful remarks.

\end{document}